\documentclass[%
 aip,rsi,amsmath,amssymb,reprint,
]{revtex4-1}

\usepackage{graphicx}% Include figure files
\usepackage{dcolumn}% Align table columns on decimal point
\usepackage{bm}% bold math

\begin{document}

\preprint{AIP/123-QED}

\title{Reservoir computing with the frequency, phase and amplitude of spin-torque nano-oscillators}

\author{D.Markovi\'c}\thanks{Author to whom correspondance should be addressed. Electronic mail : danijela.markovic@cnrs-thales.fr}
 \affiliation{Unit\'e Mixte de Physique, CNRS, Thales, Univ. Paris-Sud, Universit\'e Paris-Saclay, 91767 Palaiseau, France}
\author{N.Leroux}
\affiliation{Unit\'e Mixte de Physique, CNRS, Thales, Univ. Paris-Sud, Universit\'e Paris-Saclay, 91767 Palaiseau, France}
\author{M. Riou}
\affiliation{Unit\'e Mixte de Physique, CNRS, Thales, Univ. Paris-Sud, Universit\'e Paris-Saclay, 91767 Palaiseau, France}

\author{F. Abreu Araujo}
\affiliation{Institute of Condensed Matter and Nanosciences, UCLouvain, Place Croix du Sud 1, 1348 Louvain-la-Neuve, Belgium}

\author{J. Torrejon}
\affiliation{Unit\'e Mixte de Physique, CNRS, Thales, Univ. Paris-Sud, Universit\'e Paris-Saclay, 91767 Palaiseau, France}

\author{D. Querlioz}
\affiliation{Centre de Nanosciences et de Nanotechnologies, CNRS, Univ. Paris-Sud, Universit\'e Paris-Saclay, 91405 Orsay France}

\author{A. Fukushima}
\affiliation{National Institute of Advanced Industrial Science and Technology (AIST), Spintronics Research Center, Tsukuba, Ibaraki 305-8568, Japan}

\author{S. Yuasa}
\affiliation{National Institute of Advanced Industrial Science and Technology (AIST), Spintronics Research Center, Tsukuba, Ibaraki 305-8568, Japan}

\author{J. Trastoy}
\affiliation{Unit\'e Mixte de Physique, CNRS, Thales, Univ. Paris-Sud, Universit\'e Paris-Saclay, 91767 Palaiseau, France}

\author{P. Bortolotti}
\affiliation{Unit\'e Mixte de Physique, CNRS, Thales, Univ. Paris-Sud, Universit\'e Paris-Saclay, 91767 Palaiseau, France}

\author{J. Grollier}
\affiliation{Unit\'e Mixte de Physique, CNRS, Thales, Univ. Paris-Sud, Universit\'e Paris-Saclay, 91767 Palaiseau, France}

\begin{abstract}
Spin-torque nano-oscillators can emulate neurons at the nanoscale. Recent works show that the non-linearity of their oscillation amplitude can be leveraged to achieve waveform classification for an input signal encoded in the amplitude of the input voltage. Here we show that the frequency and the phase of the oscillator can also be used to recognize waveforms. For this purpose, we phase-lock the oscillator to the input waveform, which carries information in its modulated frequency. In this way we considerably decrease amplitude, phase and frequency noise. We show that this method allows classifying sine and square waveforms with an accuracy above 99\% when decoding the output from the oscillator amplitude, phase or frequency. We find that recognition rates are directly related to the noise and non-linearity of each variable.  These results prove that spin-torque nano-oscillators offer an interesting platform to implement different computing schemes leveraging their rich dynamical features.
\end{abstract}

\maketitle

Spin-torque nano-oscillators are promising for neuromorphic computing \cite{Macia2011, Pufall2015, Nikonov2015, Yogendra2015, Grollier2016}. These magnetic tunnel junctions can indeed emulate important properties of artificial neurons through the non-linearity and relaxation properties of current-induced magnetization dynamics. It has been shown recently that a time-multiplexed, single oscillating junction can enable or improve classification of different waveforms, distinguishing sines from squares, and even spoken digits \cite{Riou2017, Torrejon2017}. In these experiments, the input waveform was encoded in the amplitude of the input voltage and the quantity used for computing was the amplitude of voltage oscillations across the junction. Other dynamical variables can potentially be leveraged for computing, such as the frequency or the phase of the oscillators, offering a compelling platform to implement different neuromorphic computing approaches. However, all of these variables tend to be highly noisy \cite{Quinsat2010, Keller2010, Grimaldi2014}, which has been shown to be detrimental to pattern classification \citet{Torrejon2017}. Magnetization dynamics indeed takes place in nanoscale magnetic volumes, which makes them sensitive to thermal fluctuations. In addition, phase noise is enhanced by amplitude noise due to the inherent coupling between the phase and amplitude of magnetization oscillations \cite{Slavin2009}. In this work, we show that these issues can be circumvented by working in a regime where the oscillator is synchronized to the input waveform that it has to process which considerably reduces magnetization fluctuations \cite{Lebrun2015}. For this purpose, we use a sinusoidal input waveform, that carries information encoded in its modulated frequency, chosen close to the spin-torque oscillator frequency. 

\begin{figure}[!h]
\includegraphics[scale=0.55]{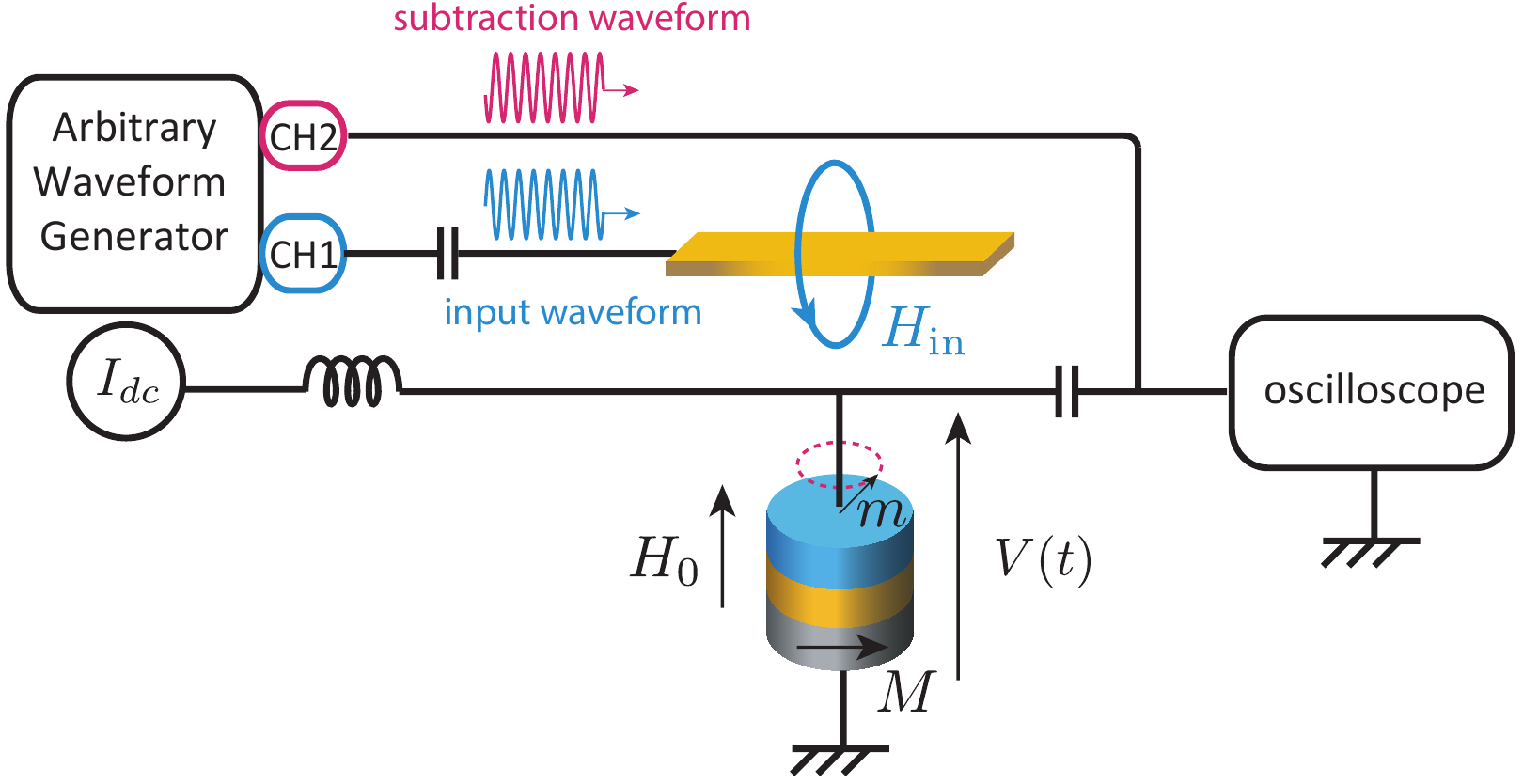}
\caption{\label{Figure1} Schematic of the measurement set-up. The spin-torque nano-oscillator is composed of two magnetic layers, of fixed magnetization $M$ (gray) and free magnetization $m$ (blue), separated by a thin insulating layer. At an external magnetic field of $H_0 = 2000$ Oe, a direct current $I_{dc} = 5$ mA is injected in order to induce magnetization precessions. The microwave signal encoding the input data in its frequency (blue) is injected into a strip line above the oscillator, thus generating a microwave magnetic field interacting with the free layer. The microwave voltage $V(t)$ emitted by the oscillator is added to a microwave signal (subtraction waveform) that compensates for the residual input signal and then is measured with an oscilloscope.}
\end{figure}

\begin{figure*}
\includegraphics[scale=0.8]{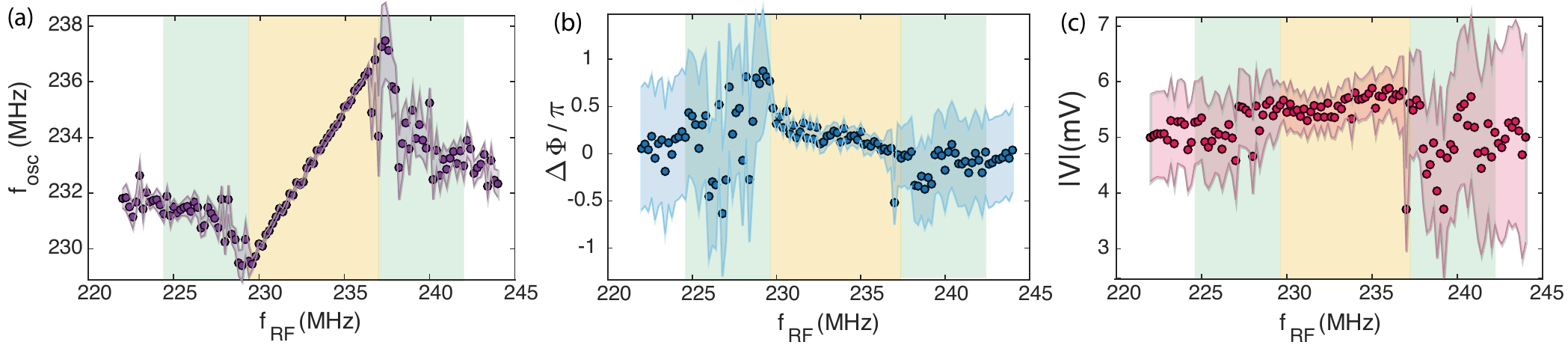}
\caption{\label{Figure2} (a) Frequency f$_{\textrm{osc}}$, (b) phase $\Delta \Phi$ and (c) Amplitude $|V|$ of the oscillator as a function of the frequency f$_{\textrm{RF}}$ of the injected microwave signal. The phase is determined with respect to that of the input waveform. Measurement uncertainties, determined on 5 $\mu$s time intervals on which the mean is calculated, are shown in lighter color shaded area. Yellow and green shaded areas designate respectively the locking range and the frequency pulling range.}
\end{figure*}

We first explain in detail our computing strategy and describe the experimental set-up used to implement it. We then show experimentally that sine and square waveforms can be classified by exploiting the frequency, phase or amplitude of the oscillations. We highlight the correlation between recognition rates and the non-linearity of these dynamical variables as a function of the input signal to classify. Our work shows that it is possible to take full advantage of magnetization dynamics by computing through all the dynamical variables describing a spin-torque nano-oscillator. In addition, since the input waveform and the oscillator output are sinusoidal waveforms with close frequencies, our scheme should allow chain-connecting the oscillators to build large neural networks. 

 Spin-torque nano-oscillators \cite{Kiselev2003, Rippard2004} are composed of two ferromagnets separated by a thin non-magnetic layer. The magnetization of the bottom ferromagnet is pinned whereas that of the top one is free. The spin-torque oscillator used in this experiment is a nano-pillar of 350 nm diameter, composed of a 1.6 nm thick CoFeB layer with a pinned magnetization, a 1 nm thick MgO insulating barrier, and a 4 nm thick FeB layer whose ground state is a magnetic vortex. Such nano-pillars can be fabricated with diameters down to 10 nm \cite{Sato2014}, which is adapted for building large scale neural networks. When a direct current is injected into this magnetic tunnel junction in  the  presence of  a magnetic  field  perpendicular  to  the  layers  stack, it induces magnetization precessions in the free layer through the effect of spin torque. Magnetoresistance effects convert magnetic oscillations into resistance oscillations, such that a microwave voltage is emitted by the oscillator and can be detected using an oscilloscope. The experimental set-up is shown in Figure~\ref{Figure1}.

Spin-torque nano-oscillators have the ability to synchronize their voltage oscillations to external microwave signals at frequencies close to their natural frequency \cite{Rippard2005, Quinsat2011, Houshang2016, Romera2017}.  The frequency and phase of the oscillator lock to the external signal frequency, and its amplitude is modified. Importantly, noise is reduced in frequency and phase so that these variables are well defined in this regime. We choose to work in this regime where the input signal synchronizes the oscillator in order to reduce the noise. The range of input microwave frequencies that synchronize the oscillator is called the injection locking range. In the following we also take advantage of the frequency pulling regime, by setting the frequency of the input signal just outside of the locking range, such that the oscillator does not get phase locked, but its frequency gets pulled towards the frequency of the input signal.

We apply a perpendicular magnetic field $H = 2000$ Oe to the oscillator and inject a direct current $I_{dc} = 5 $ mA, which induces voltage oscillations of amplitude $|V| = 13$ mV at a frequency of 232.1 MHz. We choose these field and current bias parameters in order to have a large locking range which is important for the frequency encoding and to minimize the linewidth and maximize the output signal. We use an Arbitrary Waveform Generator (AWG) to generate microwave signals that we inject into a strip line patterned 350 nm above the oscillator. This signal induces a microwave magnetic field on the oscillator, as well as a microwave current in the oscillator due to capacitive coupling with the strip line. In order to synchronize the oscillator, amplitudes of $\approx 350$ mV of the injected signal need to be applied, such that the total voltage detected by the oscilloscope is dominated by a residual capacitive microwave tone rather then the oscillator voltage. We compensate for this residual tone by adding the output voltage in a power combiner to an exactly opposed microwave signal waveform (subtraction signal in Figure~\ref{Figure1}) delayed by the time $t_0$ that the input signal takes to travel through the lines and that we calibrate prior to the measurement.

 In order to characterize the synchronization of the spin-torque oscillator with an external source, we send 5 $\mu$s long waveforms modulated at different frequencies in a 20 MHz range within the natural frequency of the oscillator. We apply the Hilbert transform \cite{Picinbono1997, Bianchini2010} on the detected voltage in order to extract frequency, amplitude and phase relative to that of the injected microwave signal, that we average over the entire 5 $\mu$s waveform. The oscillator frequency, phase and amplitude as a function of the frequency of the injected microwave signal are shown in Figure~\ref{Figure2}(a-c). As the injection signal frequency approaches the natural oscillator frequency, oscillator frequency first gets pulled towards it and then becomes identical to it in the locking range. Noise is reduced in all three variables in the locking range. Due to the subtraction of the residual microwave signal performed using a power combiner, the detected amplitude of the oscillator voltage is divided by two. This results in low signal-to-noise ratio even in the locking range [note large error bars in Figure \ref{Figure2}(c)]. The locking range, highlighted in yellow in Figure \ref{Figure2}, is experimentally determined from the standard deviation of the phase that strongly decreases in this range and is found to be 7.2 MHz.  As expected, the measured frequency of the oscillator is equal to the injected frequency in the locking range [Figure~\ref{Figure2}(a)]. The phase difference between the oscillator and the input signal roughly follows the arcsin dependence on the input frequency predicted by theory \cite{Slavin2009} [Figure~\ref{Figure2}(b)]. 
 
 \begin{figure}[!h]
\includegraphics[scale=0.5]{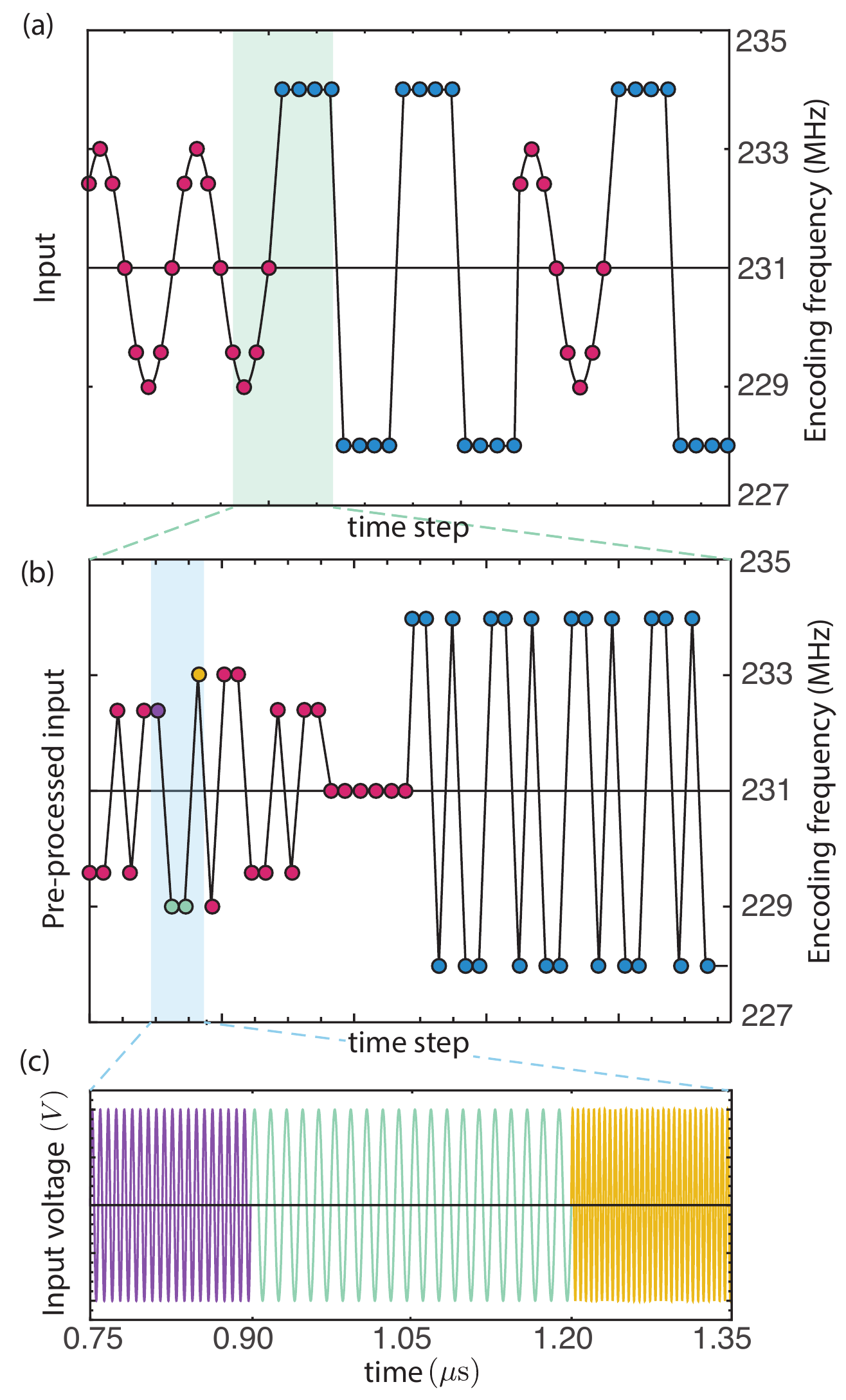}
\caption{\label{Figure3}(a) The input data is a sequence of random sine and square waves of equal periods and different amplitudes discretized in 8 points. (b) Pre-processed data corresponding to half a sine wave followed by  half a square one. In this example, the mask maps the problem to six virtual neurons. Y axis corresponds to one example of encoding frequencies.(c) Sketch of the input voltage corresponding to four neuron entries for a sine wave. Different input values are represented in different colors. The waveform amplitudes are encoded in the frequency of the microwave voltage that is then injected into the strip line for 150 ns for each data point. }
\end{figure}

 An oscillator can only achieve good performance at neuromorphic computing if it transforms the input signal in a non-linear manner \cite{Appeltant2011, Paquot2012, Torrejon2017, Larger2017}. In the pulling regime (green in Figure \ref{Figure2}), the oscillator frequency, phase and amplitude are all highly non-linear. The oscillator frequency is linear over the entire locking range, whereas the phase difference and the oscillator amplitude are non-linear at the edges [Figure~\ref{Figure2}(b)-(c)]. 

The fact that frequency, amplitude and phase are all non-linear functions of input frequency, enables us to use them to compute as with an artificial neuron. We now demonstrate this capability on a task that consists in classification of sine and square waves of equal periods but different amplitudes. For this we use a method called single node reservoir computing \cite{Appeltant2011, Paquot2012, Torrejon2017, Riou2017}. This method uses time multiplexing in order to emulate a reservoir with a single nano-oscillator that plays a role of a different effective virtual neuron at each time step.

The input data encoding procedure is shown in Figure~\ref{Figure3}. The input data is a sequence of 100 waveforms randomly chosen between sines and squares of equal frequency and different amplitudes : the amplitude of the square wave is 50\% larger than that of the sine wave. Half of this data is used for training and the other half for testing the performance. Each waveform is discretized into 8 points [see Figure~\ref{Figure3}(a)] and the task consists in determining which of the two waveform types each point belongs to. This is a non-linearly separable task and thus represents a good benchmark for neuromorphic computing \cite{Paquot2012, Torrejon2017, Riou2017}. 

Time multiplexing is achieved by preprocessing the input data as illustrated in Figure~\ref{Figure3}(b). The detailed procedure can be found in previous work \cite{Torrejon2017}. Each input point is multiplied by the same binary valued sequence called mask, whose length $N$ determines the size of the emulated reservoir. Figure~\ref{Figure3}(b) is an illustrative schematic for a reservoir containing $N = 6$ neurons, whereas in our experiment we have used $N = 25$ virtual neurons. The output of the neural network for each input time step is a weighted sum of the outputs of each virtual neuron corresponding to this input, 
\begin{equation}
\centering
y = \sum_{i=1}^{N} W_i f_{\textrm{NL}}(x_i),
\end{equation}
where $N = 25$ is the number of neurons, $W_i$ is the weight matrix element corresponding to the $i^{th}$ neuron, $f_{\textrm{NL}}$ is the non-linear function implemented by the nanodevice and $x_i$ is the input of the $i^{th}$ neuron, that is the corresponding microwave frequency. The weight matrix is calculated on a computer in order to match the target $\widetilde{y}$ = 0 or 1 respectively for sines or squares. For a target vector $\widetilde{Y}$ containing targets $\widetilde{y}$ for all the training examples, the weight matrix is calculated as $W = \widetilde{Y}F^\dagger$, where $F^\dagger$ is the Moore-Penrose pseudo-inverse of the matrix $F$ containing outputs $f_{\textrm{NL}}(x_i)$ of all neurons and for all training examples \cite{Appeltant2011}.

%%\section{Results}

\begin{figure*}
\includegraphics[scale=0.8]{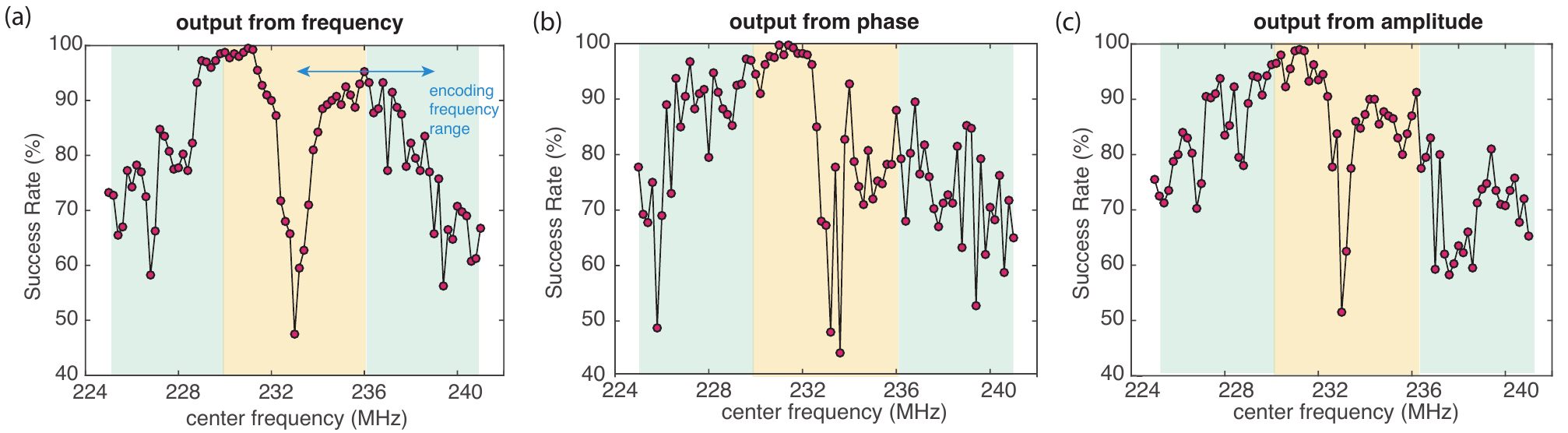}
\caption{\label{Figure4} Success rates obtained when decoding from frequency, phase and amplitude of the oscillator, as a function of the center of the frequency range chosen for encoding the input data. The frequency range used for encoding is indicated by a blue double arrow for two measurement points. Yellow and green shaded areas designate respectively the locking range and the frequency pulling range.}
\end{figure*}

Classification performance is highly dependent on the frequency window chosen for input data encoding. We choose this window in a partial or full overlap with the oscillator locking range. We fix the window width such that sine and square waves always take values in a range of 4 MHz and 6 MHz respectively. We repeat the encoding and measurement procedure for center frequencies of the encoding window varied between 225 MHz and 241 MHz and we calculate the success rate. Recognition rates obtained when decoding neuron outputs from frequency, phase and amplitude are shown in Figure~\ref{Figure4}((a)-(c)) as a function of the center frequency of the sliding window. 

Changing the center frequency has a double impact on output variables, that is the presence of noise, and the non-linear dependence on the input frequency. Noise is minimized in the middle of the locking range but the output in this regime is a linear function of the input [see Figure 2], which results in a disability of the neural network to learn the task. Indeed, as can be seen in Figure~\ref{Figure4}((a)-(c)), success rate for the frequency encoding window centered in the middle of the locking range is 50\% for all the three output variables, which for this task corresponds to random choice. The linear regime is larger for frequency than for amplitude and phase, which is reflected in the bad performance for a larger number of center frequencies in the middle of the locking range.

We find the best performance for the center frequency on the edge of the locking range, with some of the frequencies used for encoding laying in the highly non-linear frequency pulling regime. The best recognition rates are obtained when neuron outputs are decoded from the phase of the oscillations (99.75\%, Figure~4(b)), as phase is both more non-linear than frequency (best recognition rate of 99.5\%, Figure~4(a)) and less noisy then amplitude (best recognition rate of 99\%, Figure~4(c)).

These high classification rates have been obtained by using relatively large input microwave amplitudes to drive the oscillator and reduce its noise. In this regime, the magnetization relaxation time, which decreases with excitation amplitude in the locking range \cite{Rippard2013}, is very short, smaller than 4 ns in our case. Therefore, the emulated neural network performs best at tasks that do not require a memory, such as classification of different inputs. When the waveforms to separate have identical input values that can only be recognized by keeping memory of past inputs, as is the case for sine and square waves with the same amplitude, the network performance is lower (82\% recognition rate at maximum). In the future, it will be interesting to study the network intrinsic memory as a function of drive amplitude and oscillator noise. In addition, an external memory can be added to the system by using a time-delayed feedback loop and re-injecting the signal emitted by the oscillator together with the input data \cite{Appeltant2011, Paquot2012, Larger2017}.

%%\section{Conclusion}

As a conclusion, we have shown that spin-torque nano-oscillators synchronized to microwave signals can emulate artificial neural networks. The frequency, phase and amplitude of the voltage emitted by the oscillator are all non-linear functions of the frequency of the input microwave signal and can be used as outputs of the network. Working with synchronized neurons has the advantage of decreasing the frequency and phase noise, which will be of particular importance when scaling down the size of nano-pillars. In addition, frequency encoding is a simple way to use the output of an oscillator to drive the input of another, thus paving the path towards neural networks composed of chain-connected spin-torque nano-oscillators.

This work was supported by the European Research Council ERC under Grant bioSPINspired 682955 and the French ministry of defense (DGA).

\end{document}